\documentclass[aps,prl,reprint]{revtex4-1}
\usepackage{blindtext,graphicx,amsmath,amssymb}

\newcommand{\beq}{\begin{equation}}
\newcommand{\eeq}{\end{equation}}

\newcommand{\that}{ {\hat{t}} } 
\newcommand{\zhat}{ {\hat{z}} }
\newcommand{\qhat}{ {\hat{q}} }

\renewcommand{\phi}{\varphi}

\begin{document}

\title{Gravitational Pulse Wave Scattering and the Geroch group}
\author{Robert F. Penna}
\email{rpenna@ias.edu}
\affiliation{Institute for Advanced Study\\ Einstein Drive, Princeton, NJ 08540 USA}

\begin{abstract}

Cylindrical gravitational pulse waves are cylindrical gravitational waves with a pulse profile in the radial direction.  
The dynamics of cylindrical gravitational pulse waves is governed by a two dimensional integrable sigma model with an infinite dimensional hidden symmetry called the Geroch group.
Piran, Safier, and Katz obtained an exact solution for a single pulse wave by making a double analytic continuation of the Kerr metric. 
Their solution describes a pulse wave that arrives from infinity, bounces off the symmetry axis, and returns to infinity.   
In this note, we obtain an exact solution for a double pulse wave by making a double analytic continuation of the double Kerr metric.  
The new solution describes a pair of pulse waves that arrive from infinity, scatter through each other, and return to infinity.  
The pulses pass through each other and emerge with their shapes intact, as in ordinary soliton scattering.  
Unlike ordinary soliton scattering, there is no time delay.  
The metric is free of the conical singularity that appears in the double Kerr metric. 
In the future, it would be interesting to understand how the Geroch group manifests in the quantum version of gravitational pulse wave scattering.

\end{abstract}

\maketitle

\section{Introduction}

The dimensional reduction of general relativity to two spacetime dimensions is an integrable sigma model with an infinite dimensional hidden symmetry called the Geroch group \cite{geroch1971method,geroch1972method,breitenlohner1987geroch,nicolai1991two}. 
Four dimensional vacuum metrics with two commuting Killing vectors can be realized as solutions of this two dimensional sigma model.

One class of solutions that can be realized this way are stationary axisymmetric black holes.  
In this case, the integrable sigma model lives on a two dimensional Euclidean half-plane with coordinates $\rho \sim r \sin\theta  > 0$ and $z \sim r \cos \theta$.  The Kerr black hole corresponds to a single soliton of the sigma model.  The double Kerr black hole corresponds to a double soliton \cite{kramer1980superposition,tomimatsu1981multi,dietz1985two}.  

Another class of solutions that can be realized this way are cylindrical gravitational waves.  The sigma model lives on a two dimensional Lorentzian half plane with coordinates $\rho >0$ and time, $t$.  Long ago, Piran, Safier, and Katz \cite{piran1986cylindrical} made the remarkable observation that an exact solution for a cylindrical pulse wave can be obtained by a double analytic continuation  of the Kerr metric that interchanges the $t$ and $z$ directions.  The new solution describes a cylindrical gravitational wave with a pulse profile in the radial direction.  The pulse arrives from infinity, bounces off $\rho=0$, and returns to infinity.  The pulse maintains its shape as it travels \footnote{Actually, the shape of the pulse steepens as it approaches $\rho=0$ because it travels on a half-line.}, as expected for a soliton.

In this note, we obtain a double pulse wave solution by making a double analytic continuation of the double Kerr metric.  The new metric describes a pair of cylindrical pulse waves that arrive from infinity, scatter, and return to infinity.  The pulses pass through each other and emerge with their shapes intact, so their soliton nature is confirmed.  Unlike ordinary solitons, they suffer no time delay after scattering.   The conical singularity of the double Kerr metric is not present in the new metric.  So the double pulse wave is, in some sense, a more physical solution.  One can contemplate the generalization to $N$ pulse waves but we will not do so here.

Earlier works \cite{carr1983soliton,ibanez1983soliton,ibaez1985multisoliton,boyd1991properties} have studied gravitational soliton scattering in cosmological spacetimes.  In those solutions, the solitons emerge from a ``big bang'' singularity at $t=0$.  An interesting feature of the solution described in the present work is that the pulses emerge from (and return to) asymptotic infinity.   Related works \cite{piran1985general,tomimatsu1989gravitational,tomizawa2015nonlinear} on gravitational soliton scattering  have studied the gravitational analogue of Faraday rotation.

An interesting challenge for the future is to understand the quantum version of gravitational pulse wave scattering.  The Poisson brackets of Geroch group charges generate a Yangian symmetry \cite{korotkin1998yangian}.   In other quantum integrable  sigma models, Yangian symmetry manifests through the absence of particle production and the factorization of the $S$ matrix \cite{zamolodchikov1978relativistic,zamolodchikov1990factorized,luscher1978quantum}.  It would be interesting to understand how the Yangian symmetry of dimensionally reduced general relativity manifests in the quantum version of gravitational pulse wave scattering.

General relativity with a negative cosmological constant admits cylindrical black hole solutions \cite{lemos1995three}.  So it might be possible to find integrable models of cylindrical pulse waves interacting with a cylindrical black hole.  Unfortunately, it is not known how to include a cosmological constant in the theory of the Geroch group (but see \cite{leigh2014geroch,Petkou:2015fvh} for work in that direction).  Perhaps the recently introduced twistor Chern-Simons theory \cite{woodhouse1988geroch,costellotalk,Bittleston:2020hfv,penna2021twistor} could offer a useful new perspective on the problem of including a cosmological constant in the theory of the Geroch group.

It would be worthwhile to perform an asymptotic symmetry analysis of the pulse wave metrics described herein along the lines of \cite{ashtekar1997asymptotic,ashtekar1997behavior,barnich2007classical}.  It would also be interesting to study the effect of small deviations from cylindrical symmetry on the dynamics.  In that case the Geroch symmetry will be weakly broken, but even weakly broken Geroch symmetry might be useful for organizing the dynamics, in the spirit of effective field theory.

\section{Result}

The double Kerr metric has the form \cite{kramer1980superposition,tomimatsu1981multi,dietz1985two} 
\beq\label{eq:metric}
ds^2 = -f (d\that - \Omega d\phi)^2 
	+ f^{-1} \left( e^{2\gamma} (d\rho^2 + d\zhat^2) + \rho^2 d\phi^2 \right) .
\eeq
There are three functions in the metric:  $f$, $\Omega$, and $\gamma$.  They are functions of $\rho$ and $\zhat$ only.   

$\rho$ is a radial coordinate ($\rho>0$) and $\zhat$ is the vertical coordinate ($-\infty < \zhat < \infty$).  The metric has two commuting Killing vectors, $\partial_\that$ and $\partial_\phi$.  $\that$ is the time coordinate.

It is convenient to replace $\Omega$ with a new function, $\psi$, defined by
\beq
\partial_\rho \psi = \frac{f^2}{\rho} \partial_\zhat \Omega \,, \quad 
\partial_\zhat \psi = - \frac{f^2}{\rho} \partial_\rho \Omega \,. 
\eeq

The vacuum Einstein equations give
\begin{align}
\partial_\rho \gamma	&=  \frac{\rho}{4f^2} \left[ (\partial_\rho f)^2 - (\partial_\zhat f)^2 
					+  (\partial_\rho \psi)^2 - (\partial_\zhat \psi)^2 \right] , \label{eq:gammarho} \\
\partial_\zhat \gamma 	&= \frac{\rho}{2f^2}\left[ (\partial_\rho f)(\partial_\zhat f) + (\partial_\rho \psi)(\partial_\zhat \psi) \right] . \label{eq:gammaz}
\end{align}
We can solve these equations for $\gamma$ once we know $f$ and $\psi$.  Actually, these equations only fix $\gamma$ up to a constant.  Fix this constant by setting $\gamma=0$ at infinity.

Combine $f$ and $\psi$ into a single complex scalar field, the Ernest potential,
\beq
\epsilon = f + i \psi \,.
\eeq
The Ernest potential for the double Kerr metric is
\beq
\epsilon = \frac{1-\xi}{1+\xi} \,,
\eeq
where
\beq\label{eq:xi}
\xi = 
\frac{
\begin{vmatrix}
1 & 1 & 1 & 1 \\
s_1 & s_2 & s_3 & s_4 \\
k_1 & k_2 & k_3 & k_4 \\
k_1^2 & k_2^2 & k_3^2 & k_4^2 
\end{vmatrix}
}
{
\begin{vmatrix}
1 & 1 & 1 & 1 \\
s_1 & s_2 & s_3 & s_4 \\
k_1 & k_2 & k_3 & k_4 \\
k_1 s_1 & k_2 s_2  & k_3 s_3 & k_4 s_4 
\end{vmatrix}
} \,.
\eeq
We recover $f$ and $\psi$ as the real and imaginary parts of the Ernst potential. 

\begin{figure}
\includegraphics[width=\linewidth]{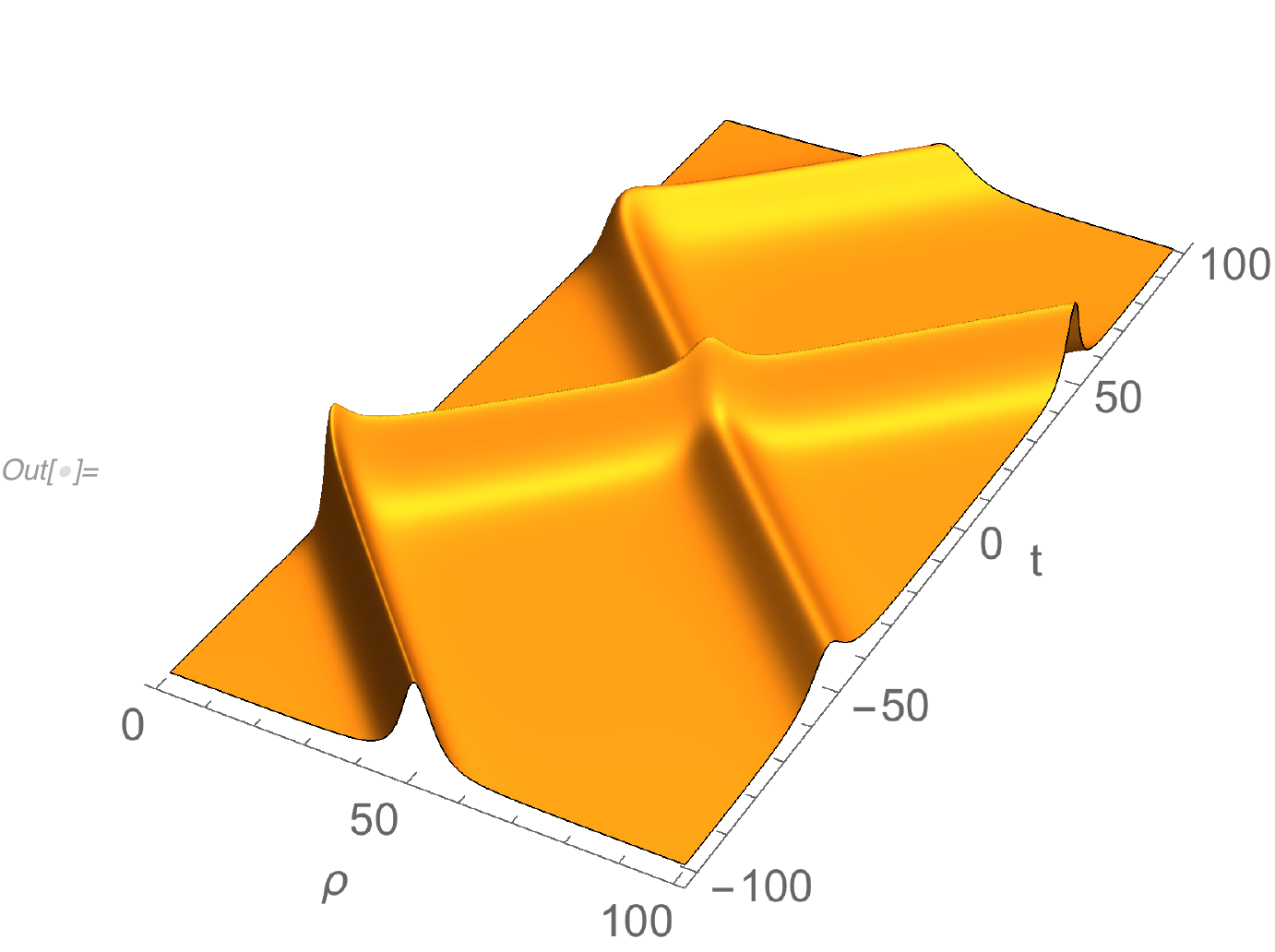}
\caption{\label{fig:plot3d} Gravitational pulse wave scattering with $m_1 = 4$, $m_2 = 2$, $q_1= q_2  = 2$ and $t_1 = -t_2 = 50$. The function plotted is $\partial_\rho \gamma$, which is related to the energy \cite{thorne1965energy}.  The $\phi$ and $z$ directions are not shown. }
\end{figure}

The numerator and denominator of $\xi$ are determinants of $4\times 4$ matrices.   The $k_m$ ($m=1,\dots,4$) in equation \eqref{eq:xi} are four real constants.  The spacetime dependence of the Ernest potential enters through
\beq
s_m = r_m e^{i \omega_m} \,,
\eeq
where the $\omega_m$ ($m=1,\dots,4$) are an additional four real constants and
\beq
r_m = \left(\rho^2 + (\zhat-k_m)^2 \right)^{1/2} \,.
\eeq
So the metric depends on $\rho$ and $\zhat$ only through $r_m$.  The metric has a total of eight real parameters, $k_m$ and $\omega_m$ ($m=1,\dots,4$).

To get a more familiar set of parameters, let
\begin{align}
k_1 &= \zhat_1 + m_1 p_1 \,, \\
k_2 &= \zhat_1 - m_1 p_1 \,, \\
k_3 &= \zhat_2 + m_2 p_2 \,, \\
k_4 &= \zhat_2 - m_2 p_2 \,,
\end{align}
and
\begin{align}
\omega_1 &= \alpha_1 + \lambda_1 \,, \\
\omega_2 &= \alpha_1 - \lambda_1 + \pi \,, \\
\omega_3 &= \alpha_2 + \lambda_2 \,, \\
\omega_4 &= \alpha_2 - \lambda_2 + \pi \,,
\end{align}
where
\begin{align}
\lambda_i  &= \frac{1}{i} \log (p_i + i \qhat_i) \,, \\
p_i  &= \sqrt{1 - \qhat_i^2 } \,.
\end{align}
The new set of eight parameters is $m_i$, $\qhat_i$, $\alpha_i$, and $\zhat_i$ ($i=1,2$).  When the black holes are widely separated, $m_i$, $\qhat_i$, $\alpha_i$ and $\zhat_i$ are, respectively, the masses, dimensionless spin parameters, NUT parameters, and vertical displacements of the holes.  We set the NUT parameters to zero from now on ($\alpha_1 = \alpha_2 = 0$).

To get the double pulse wave metric, we make the double analytic continuation
\beq\label{eq:wick}
\that = i z \,, \quad \zhat = i t \,,
\eeq
where $t$ and $z$ are real coordinates.  To get a real metric, we need to choose
\beq
\qhat_i = i q_i \,, \quad
\zhat_i = i t_i  \quad  (i=1,2),
\eeq
where $q_i$ and $t_i$ are real parameters.

The new metric describes a pair of pulse waves that arrive from infinity, scatter through each other, and return to infinity.   Figure \ref{fig:plot3d} depicts a solution with $m_1=4$, $m_2=2$, $q_1=q_2  = 2$, and $t_1=-t_2=50$.  The most interesting feature  is the interaction region where the pulse waves momentarily merge. The pulses emerge from the interaction region with their original shapes intact.  This is a characteristic feature of soliton scattering.

\begin{figure}
\includegraphics[width=0.6\linewidth]{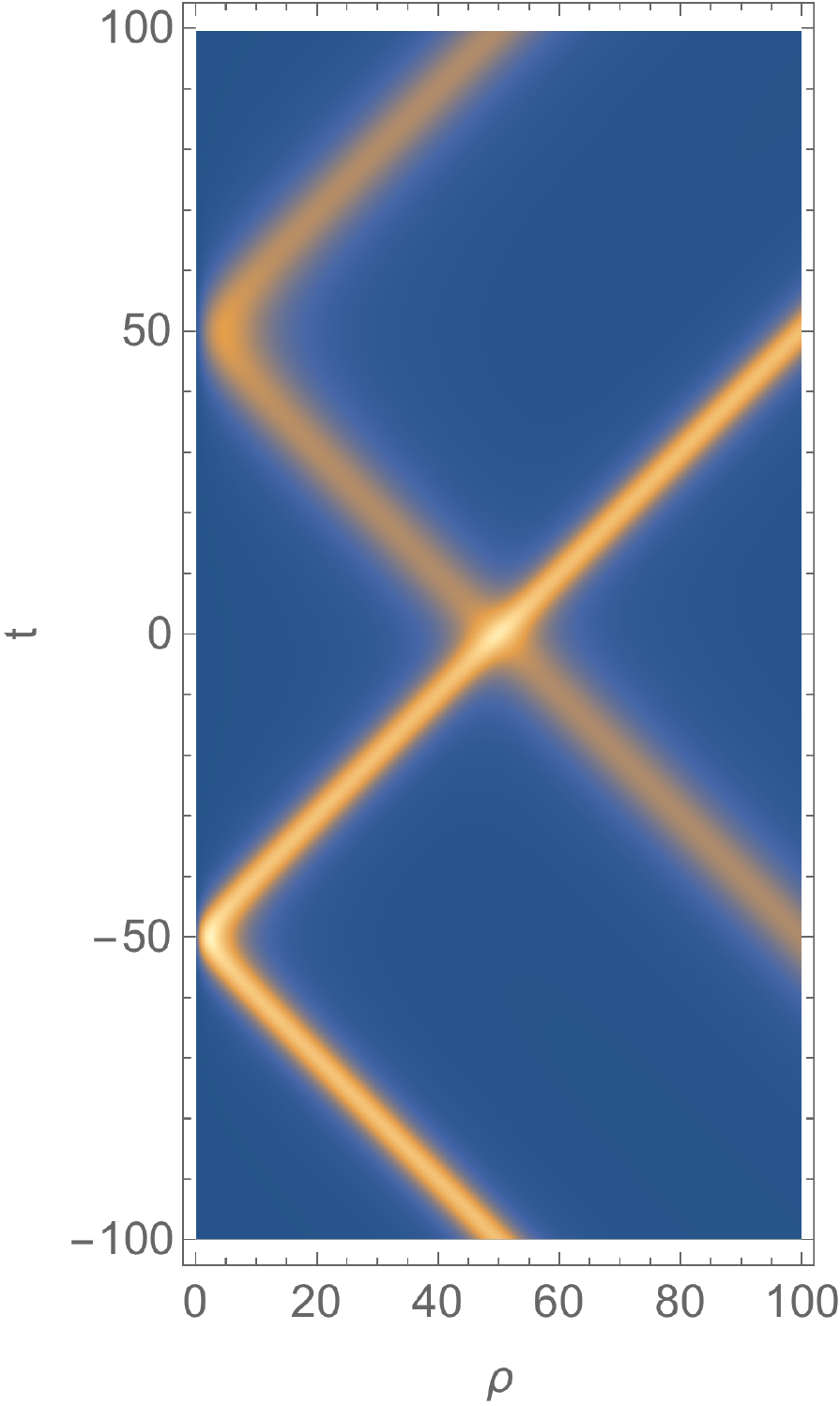}
\caption{\label{fig:densityplot} Same as Figure \ref{fig:plot3d} but viewed from above.  It is clear from this figure that there is no time delay.  The second pulse reaches $\rho=0$ at $t=50$, just as it would if there were no interaction.}
\end{figure}

It is often the case in soliton scattering that the solitons emerge from an interaction with a time delay.  However, there is no time delay in the present example.  This is illustrated in Figure \ref{fig:densityplot}.  The second pulse reaches $\rho=0$ at $t = 50$, just as it would with no scattering.  The absence of time delays for cosmological solitons was observed by \cite{boyd1991properties}.

The black holes in the  double Kerr metric are connected by a conical singularity that runs along the axis at $\rho=0$.  The conical singularity acts as a strut and holds the black holes apart.   

There is no such conical singularity in the double pulse wave metric.  To see why, consider equation \eqref{eq:gammaz} at $\rho=0$:
\beq
(g_{\that\that})^2 \partial_\zhat \gamma = 0 \,.
\eeq
It follows from this equation that $\gamma$ is a piecewise constant function of $\zhat$ at $\rho = 0$.  In the double Kerr metric, there are jumps in $\gamma$ at the ``ergosphere,'' where $(g_{\that\that})^2 = 0$.  The conical singularity appears at the jumps in $\gamma$.  

In the case of the double pulse metric, we have instead
\beq
(g_{zz})^2 \partial_t \gamma = 0 \,.
\eeq
Now $(g_{zz})^2$ is strictly positive, so there are no jumps in $\gamma$ and no conical singularity.

The single pulse wave solution of Piran, Safier, and Katz \cite{piran1986cylindrical}  requires periodic identifications in the $\phi$ and $z$ directions which are related to the thermodynamics of the Kerr black hole.  We expect the periodicity of the double pulse wave metric in the $\phi$ and $z$ directions to be related to the rather more complicated thermodynamics of the double Kerr metric \cite{herdeiro2010thermodynamical}.

\begin{center}
\textbf{ Acknowledgments}
\end{center}
I am grateful to Ibrahima Bah for helpful comments.  This material is based upon work supported by the U.S. Department of Energy, Office of Science, Office of High Energy Physics under Award Number DE-SC0009988.

\bibliographystyle{apsrev4-1} 
\bibliography{scattering} 

\end{document}